\documentclass[aps,showpacs,prl,twocolumn]{revtex4}
\usepackage{amsmath,amssymb}
\usepackage{graphics}
\usepackage{graphicx}

\begin{document}

\title{Time dynamic of Fourier modes in turbulence:\\
Sweeping effect, long-time correlations and time intermittency}
\author{C. Poulain\footnote{permanent address: \textit{Commissariat \`a l'Energie Atomique (CEA) }\\  17 rue des Martyrs, 38054 Grenoble, France}}
\author{N. Mazellier}
\author{Y. Gagne}
\author{C. Baudet}
\affiliation{ \textit{Laboratoire des Ecoulements G\'eophysiques et Industriels}\\  LEGI 38041 Grenoble, Cedex 9, France}

\date{\today}

\begin{abstract}
                                                       
We present an experimental study of the typical time scales of spatial Fourier modes of the vorticity in turbulence. The measurements of time auto-correlation functions, 
for well defined spatial wavenumbers $k$, reveal that vorticity exhibits two characteristic times: a short one depending on the inverse of the wavenumber, 
which can be interpreted as a manifestation of the sweeping effect (random advection), and a larger one, independent of $k$, of the order of the integral time scale of the flow.
In a second step, we show that two different vorticity Fourier modes with wavenumbers $k$ and $k'$ are long-time cross-correlated, whatever the spectral gap $k-k'$. 
The existence of a significant cross-correlation between scales, indicates a strong statistical dependance between scales, at variance with the original Kolmogorov theory (1941). 
This long-time correlation, closely related to the integral forcing time, could be ascribed to some evidence of time intermittency, associated with fluctuations in time of 
the energy injected at large scales and resulting in a lack of global scale invariance of the dynamical properties of the turbulent flow.
\end{abstract}


\maketitle
Statistical intermittency, responsible for a lack of statistical global scale invariance, remains one of the main puzzling features of turbulence, still unresolved \cite{Frisch1995}.
At the experimental level, following Kolmogorov initial prediction \cite{Kolm1941}, the evolution across scales of the turbulence statistical properties can be traced by 
computing the spatial velocity increments, $\delta u_r(x, t)=u(x+r, t) - u(x, t)$. 
Usually, one resorts to hot-wire anemometry providing Eulerian measurements of the longitudinal velocity component at one point $x_o$ along time $t$. 
Scale dependence is then recovered by mapping time increments $\delta t$ of the Eulerian signal onto spatial increments $r=-U_{avg}\delta t$ according to the 
Taylor hypothesis of \emph{frozen turbulence} \cite{Frisch1995}: $\delta u_r(t)=u(x_o,t) - u(x_o, t-r/U_{avg})$. 
In this context (spatial intermittency), and according to Kolmogorov \emph{refined similarity hypothesis} \cite{Frisch1995, Kolm1962}, statistical intermittency is the consequence 
of the strong spatial heterogeneity (possibly multifractality) of the local energy transfer rate, related to some kind of multiplicative cascade process of the turbulent energy across scales. 
In the spirit of Kolmogorov approach, the energy transfer rate is also expected to display strong time fluctuations, reflecting unavoidable fluctuations of the energy 
injected at large scales. Recently, it has been suggested \cite{Pandit2003}, that the study of dynamical multiscaling (time dependence of the velocity structure functions) 
could be of valuable help for a better comprehension of turbulence intermittency.
Unfortunately, as implicitely stated by the Taylor hypothesis, the statistics of Eulerian velocity increments are strongly dominated by the spatial features of the flow and thus,   
weakly sensitive to its time fluctuations. 
Actually, it is generally accepted that the latter fact is a direct consequence of the random character of the advection by the large scale flow, past any Eulerian probe, 
of the whole velocity field (the so-called sweeping effect \cite{Kraichnan1989, Nelkin1990}).

\begin{figure}
\center{\includegraphics[width=7cm]{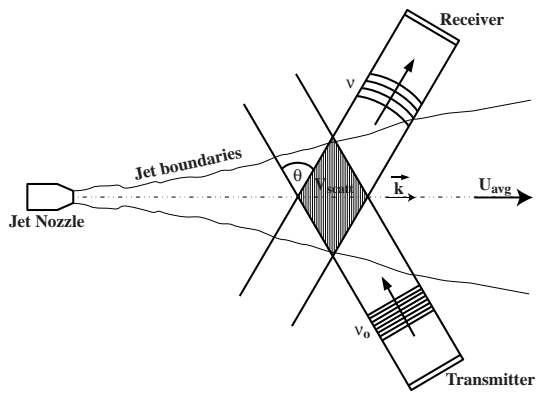}}
\caption{Sketch of the experimental setup: turbulence is produced by a turbulent round jet at $R_{\lambda} \simeq 600$. 
The acoustic measurement volume $V_{scatt}$ is located downstream the nozzle. The wavevector $\vec{k}$ is parallel to the mean flow axis ($x$) 
while the probed component of the vorticity vector field is perpendicular to the scattering plane.}
\label{fig:expsetup}
\end{figure}

An alternate and efficient way to study statistical scale dependence is to rely on a spatial Fourier analysis of the field of interest \cite{Kraichnan1977, Pandit1997}. Spectral measurements, based  
on wave scattering experiments are commonly used in various domains of research in physics, starting with phase transitions and critical phenomena in condensed matter physics 
where light and neutron scattering techniques are largely widespread. 
It is now well established on both theoretical \cite{Kraichnan1953, Lund1989} and experimental basis \cite{Baudet1991}, that acoustic waves propagating in a turbulent medium 
can be scattered by vorticity fluctuations.
As in any scattering experiments, it may be shown \cite{Lund1989} that, due to the coherent average of the waves scattered by the vorticity distribution, the overall scattered 
amplitude $p_{scatt}(t)$ is linearly related to the spatial Fourier transform of the vorticity field according to :  

\begin{equation}\label{eq:ModulationAmplitude}  
p_{scatt}(t) \propto \widetilde{\omega}_\perp\left(\vec{k},t\right).p_{0}(t)
\end{equation}

where 

\begin{equation}\label{eq:ScattAmplitude}
\widetilde{\omega}_\perp\left(\vec{k},t\right) = \iiint_{\mbox{V}_{{\small \mbox{scatt}}}}
\omega_\perp\left(\vec{x},t\right) e^{-i \vec{k}\cdot\vec{x}} d^3x
\end{equation}

the scattering wavevector $\vec{k}$ being a function of the incoming sound frequency $\nu_0$ and the scattering angle $\theta$: 
\begin{equation}\label{eq:def_k}\vec{k} = 4 \pi \nu_0\sin\left(\theta/2 \right)/c \,\vec{x}\end{equation} 
with $c$ the sound speed. Note that only the component of the vorticity vector field, normal to the scattering plane, is 
involved in the scattered amplitude (hence the index $_\perp$).
The scattering setup consists in a bistatic configuration (Fig. \ref{fig:expsetup}): an ultrasound monochromatic plane wave $p_{0}(t)$, with frequency $\nu_0$, continuously 
insonifies the turbulent media and the acoustic amplitude $p_{scatt}(t)$, scattered in the direction $\theta$, is recorded along time by a receiver.
Both acoustic transmitter and receiver work in a linear regime (they are phase sensitive). According to equation (\ref{eq:ModulationAmplitude}), a direct 
image of the spatial Fourier mode of the vorticity at wavevector $\vec{k}$, is obtained by a simple heterodyne demodulation providing a complex signal (phase and amplitude).

The main interest of the acoustic scattering experiment lies in the fact that, contrary to classical Eulerian measurements, the scale dependence (through the 
selection of a unique wavevector $\vec{k}$) is performed independently of the time evolution of the collected signal, thanks to the direct spatial 
Fourier transform involved in the scattering process.
We will now focus on the time behavior of the Fourier modes of vorticity $\widetilde{\omega}\left(\vec{k},t\right)$, as a function of the length scale parameter $k$.
Although a Fourier mode is a complex quantity (as is the demodulated scattered signal), we will restrict ourselves to the amplitude of the signal:
$|\widetilde{\omega}_\perp\left(\vec{k},t\right)|$ hereafter noted $\omega(k,t)$ for sake of simplicity.
We have investigated the statistical properties of a turbulent round air jet at room temperature. 
The direction of the probed vorticity field is radial, while the direction of the scattering wavevector $\vec{k}$ is aligned with the mean flow velocity.
The flow emerges in the $x$ direction from a nozzle of diameter $D=0.12m$ (see Fig. \ref{fig:expsetup}).
%
%
Throughout the experiment, the scattering angle is kept at a constant value and different wavevectors are analysed by varying the 
incoming sound frequency $\nu_o$. 
As the scattering angle $\theta$ is constant, one can show that, in our bistatic configuration, the spectral resolution is given by 
$\delta k\sim V^{-1/3}_{scatt}$, independent of the analysed wavenumber $k$ \cite{Poulain2004}. 
The measurement volume $V_{scatt}$ is defined by the intersection of the incident and detected acoustic beams and mainly
depends on $\theta$  and on the size of both acoustic transducers. 
With $\theta = 40^{o}$ and a diameter of the circular transducers of $14 cm$, the linear extension of $V_{scatt}$ is of the order of the integral 
length scale of the jet flow estimated to $L=0.36m$.
Additional flow parameters have also been estimated, using conventional hot-wire anemometry: the Taylor micro-scale is $\lambda_T=7.6mm$ 
and the associated Taylor-based Reynolds number $R_\lambda=u'\lambda_T/\nu$ worths $R_\lambda\simeq600$ where $u'$ is 
the longitudinal root-mean-square of the longitudinal velocity fluctuations.

\begin{figure}
\center{\includegraphics[width=6cm]{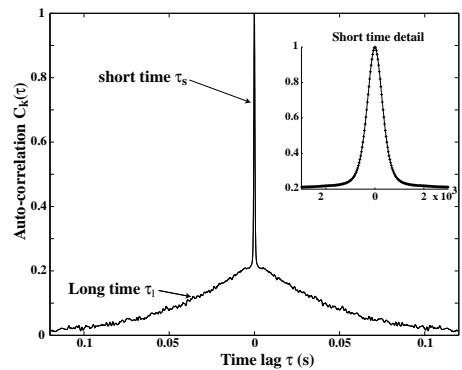}}
\caption{Time auto-correlation function $C_k (\tau)$  at wavenumber $k=1200m^{-1}$. The global shape is similar for all analysed wavenumbers: 
a short-time decorrelation with characteristic time $\tau_S$ (see insert for detail), followed by a much slower decrease, with 
(see also Fig. \ref{fig:intercloglin} (a) for a log-lin plot) and characteristic time $\tau_L$.  
Here, $\tau_S=0.88 ms$ and $\tau_L=45 ms$.
\label{fig:autoclinlin} }
\end{figure}

\begin{figure}
\center{\includegraphics[width=8cm]{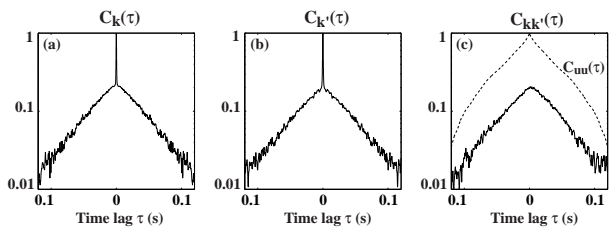}}
\caption{Log-lin plots of time correlation functions:
(a) auto-correlation $C_k(\tau)$ for $k=1200m^{-1}$ (as in Fig. \ref{fig:autoclinlin}), 
(b) auto-correlation $C_k'(\tau)$ for $k=820m^{-1}$,  
(c) solid line: cross-correlation $C_{kk'}(\tau)$ between the two foregoing Fourier modes, dotted line: auto-correlation $C_{uu}(\tau)$ of the Eulerian
longitudinal velocity $u$ at the center of $V_{scatt}$.}
\label{fig:intercloglin}
\end{figure}

From the time signals $\omega(k,t)$, collected at a fixed wavevector $\vec{k}$, we compute the time auto-correlation function 
$C_k (\tau)=<\omega(k,t)\,\omega(k,t-\tau)>_t$ where $<.>_t$ stands for the time average. 
A typical evolution of the normalized $C_k(\tau)$ ($C_k(0) = 1)$), with respect to the time lag $\tau$, is sketched on figure \ref{fig:autoclinlin}.
Whatever the turbulent scale $k$, the same global shape is found, exhibiting two different and well separated characteristic times. 
For time lags close to zero, one observes a rapid decrease, with a more or less Gaussian shape. 
At larger time lags, a much slower decrease is visible, with a nearly exponential behavior as evidenced on the semi-log representation in Figure 
\ref{fig:intercloglin} (a). 
Let us estimate the short time $\tau_S$, by measuring the half amplitude width ($C_k (\tau_S) = 1/2 $ in the small lags region, 
and the long time $\tau_L$, with an exponential fit of the tails of $C_k(\tau)$ at large times.
Although such a Gaussian shape at short times has been predicted in some theoretical models \cite{Heisenberg1948, Kraichnan1959}, the long time behaviour 
does not seem to have ever been reported. 
%
%
\begin{figure}
\center{\includegraphics[width=6cm]{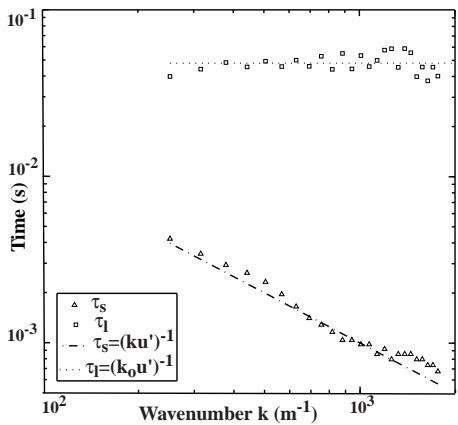}}
\caption{Loglog plot of the evolution of the two characteristic times of Fourier modes in turbulence. 
\label{fig:tauStauL} }
\end{figure}

We now turn to the scaling of those two characteristic times with the wavenumber $k$. 
By tuning the incoming sound frequency ($10 kHz\le\nu_o\le170k Hz$), we have successively probed a decade of wavenumbers $k$, spanning the inertial 
range of the turbulent flow down to the dissipative range ($0.30 < k\lambda_T/2\pi < 4.7$).
Figure \ref{fig:tauStauL} clearly reveals two different behaviours: $\tau_S$ is scale dependent while $\tau_L$ is not.  
Systematic studies, performed in various flow configurations (velocity $u'$ and integral scale $L$) gave us reliable results for both times. 
As for the short time $\tau_S$, it decreases with $k$ following the power law $\tau_S\propto (k u')^{-1}$. 
In our experiments, the proportionality constant is about 1, and slightly decreases with the Reynolds number.  
Such a scaling law, involving the root mean square velocity (a large scale quantity), is usually ascribed to a sweeping effect associated with 
the random advection of the vorticity field by the large scale velocity fluctuations, according to Navier-Stokes equation.
The consequence of the sweeping, with respect to Eulerian turbulence statistics, is a controversial question as regards the $-5/3$ exponent of the 
power-law scaling of the turbulent energy-spectrum \cite{Kraichnan1989, Nelkin1990, Tennekes1975, Yakhot1989, Praskovsky1993}.
However, it is generally well accepted that the sweeping effect is responsible for the unexpected (according to the global scale invariance hypothesis) 
$f^{-5/3}$ power law behaviors of the spectra of powers ($p>2$) of the Eulerian turbulent velocity fluctuations $|u'|^p$ (where $f$ is the temporal frequency) \cite{Nelkin1990}.
Accordingly, our experimental results confirm that this latter effect can be more clearly evidenced in the Fourier domain as suggested in reference \cite{Gorman2004}.
Let us now turn to the large time $\tau_L$: as we have checked extensively (by varying acoustic as well as hydrodynamic conditions separately) $\tau_L$ is 
\emph{not} scale dependent but rather depends on the forcing scales. Actually, we found that:  $\tau_L\propto (k_L u')^{-1}$ where $k_L=2\pi/L$ is the large scale wavenumber.
The fact that this long time correlation is not scale dependent and, besides, remains almost identical whatever the probed scale, 
with a significant correlation level (about 20\% Figs.  \ref{fig:intercloglin} (a)-(b)) is noteworthy.
First, the time $\tau_L$ is exactly proportional to the integral Eulerian time (extracted from the velocity auto-correlation fonction see Fig. \ref{fig:intercloglin} (c)), 
with a proportionality constant of one: $\tau_L\approx (k_L u')^{-1}$. 
Second, the long-time invariant behaviour through the scales, contrary to the scale dependence of the short time, indicates that it is not possible to 
find a scaling transformation which leads to a collapse of all the correlation functions onto a single one.  
We are lead to the conclusion that the coexistence of these two characteristic times indicates a lack of global scale invariance in the cascade process. 
In the spirit of spatial intermittency, related to the absence of an universal shape of the probability density functions of the spatial velocity increments, 
it is thus tempting to term our observations as a manifestation of time intermittency. 
It is worth mentionning, at this point, that we have also evidenced some influence of the spatial intermittency of turbulence, manifested by a significant dependence  
of the statistics of the spatial Fourier modes on the spectral resolution $\delta k$ or the spatial extension $V^{-1/3}_{scatt}$ of the experimental 
setup with respect to the integral length scale of the turbulent flow \cite{Chevillard2005}.

The close resemblance in the long time evolution of any pair of spatial Fourier modes with different wavenumbers $k$ and $k'$, suggests that they could be 
dynamically driven by the very same large scale process. 
To get further insight about this observation, we have performed simultaneous acquisitions of two distinct spatial Fourier modes $k$ and $k'$.
Provided the spectral gap $k-k'$ is large enough, such measurements can be performed by driving a single transmitter with the sum of two sine waves with 
the appropriate frequencies $\nu_o$ and $\nu_o'$ (Eqn. \ref{eq:def_k}).  
The scattered pressure signals, around each incoming frequency, can be easily separated by means of two simple band-pass filtering operations.
To avoid spurious interference effects, the same investigation could also be performed by using a second pair of transducers defining a second independent 
scattering channel as in \cite{Baudet1999}. 
However, the single pair configuration presents the advantages of a better wavevector alignment as well as the best possible measurement volume matching. 
Actually, we have carefully checked that both setups give the same results. 
From two \emph{synchronous} time series (at $k$ and $k'$), with a well controled spectral gap $k-k'$,  we have computed the cross-correlation function 
$C_{k,k'}(\tau)=<\omega(k,t)\,\omega(k',t-\tau)>_t$, of two spatial Fourier modes. 
A typical example of this cross-correlation is represented on Fig. \ref{fig:intercloglin} (c) (the same behaviour is
observed whatever the scale separation). We have also represented on Fig. \ref{fig:intercloglin} (a) (resp (b)), the auto-correlation function of the two spatial 
Fourier modes with wavenumbers $k$ (resp. $k'$).
A significant level of cross-correlation is observed for time lags up to the integral time scale, with a shape and an amplitude close to that of both auto-correlation functions. 
The main difference between auto-correlation and cross-correlation functions lies in the absence of the rapid decay at small time lags for the cross-correlation.
We are lead to the conclusion that the two observed behaviours (short and long time correlation) should be associated to two different dynamical process driving the vorticity field. 
We also point out the fact that the existence of a significant cross-correlation between the two spatial Fourier modes $k$ and $k'$ implies a strong statistical 
dependance between scales, irrespective of the spectral gap $k-k'$. 
This new result, is clearly at variance with the initial Kolmogorov 1941 theory \cite{Kolm1941}.  
As far as the long time driving process is concerned, the significant cross-correlation level suggests that all scales are \emph{instantaneously} driven 
by  the same process (possibly multiplicative).
Our belief is that the observed long time statistical dependance could be related to some memory effect of the time fluctuations of the energy 
injection at large scales, in the spirit of Landau's objection formulated against the K41 model in 1944 \cite{Frisch1995, Landau1987}.

To summarize, acoustic scattering allows the direct spectral probing, continuously in time, of spatial Fourier modes of one component of the vorticity vector field. 
Thanks to a good spectral resolution, the proper selection of a well defined wavevector $\vec{k}$ of the turbulent flow results in an unambiguous separation of the 
spatial and time features of the turbulent dynamic. 
As a first result, we clearly evidence sweeping effects, related to the randomness of the large scale velocity field advecting the small scales and manifesting 
as a rapid decorrelation with a nearly Gaussian shape (reflecting the Gaussian statistics of the spatio-temporal advecting velocity field). 
A much slower modal decorrelation, over time lags up to the integral time scale, is also  observed and reveals a significant statistical dependency between scales.
We propose to ascribe our experimental observations to some large scale time intermittency (in contrast with the usual small scale spatial intermittency affecting 
the statistics of Eulerian spatial velocity increments). 
A plausible explanation for the source of such time intermittency, could be the random nature of the time fluctuations of the rate of energy injected at
large scales. 
To conclude, we want to stress that, as we have observed similar statistical behaviours at smaller Reynolds numbers (in a grid turbulence at $Re_\lambda \simeq 100$), 
as well as at much higher Reynolds numbers (up to $Re_\lambda \simeq 6000$, in a cryogenic Helium jet facility at CERN \cite{GReC2002}), 
this could well be a generic feature of flow turbulence.

This work is fully supported by the French Minist$\grave{\mbox{e}}$re de la Recherche and Universit\'e Joseph Fourier (PPF plateforme
exp\'erimentale de spectroscopie acoustique multi-\'echelles). We wish to acknowledge Laurent Chevillard for fruitful discussions. 
We also thank Jean-Paul Barbier-Neyret and Joseph Virone for their valuable technical help.

\end{document}